\documentclass[10pt, twocolumn]{LML}

\newcommand{\sa}{\braket{\sigma_{ann} v}}
\newcommand{\ssa}{\braket{\sigma_{sem} v}}
\begin{document}


\title{Impact of loop-induced processes on the boosted dark matter interpretation of the {\tt XENON1T} excess}

\author[a]{Luigi Delle Rose\thanks{luigi.dellerose@fi.infn.it}}
\author[b]{Gert H\"utsi\thanks{gert.hutsi@to.ee}}
\author[b]{Carlo Marzo\thanks{carlo.marzo@kbfi.ee}}
\author[b]{Luca Marzola\thanks{luca.marzola@cern.ch}}

\affil[a]{INFN, Sezione di Firenze and Department of Physics and Astronomy, University of Florence,
Via G. Sansone 1, 50019 Sesto Fiorentino, Italy.}
\affil[b]{\KBFI}
\date{\today}

\twocolumn[
\maketitle
\begin{onecolabstract}
	We consider the boosted dark matter solution of the {\tt XENON1T} excess to constrain the framework through loop-generated processes. The interaction of the boosted dark matter component, which sources the signal, effectively couples the cold dark matter background to the electrons, making it potentially visible in the electron recoil searches. Similarly, once the radiative corrections due to the Standard Model are taken into account, dark matter also scatters on quarks and becomes observable in nuclear recoil measurements. By analysing these processes, we find that the current direct detection constraints exclude the upper mass range selected by the anomaly if the boosted component is generated through dark matter annihilation.    
\end{onecolabstract}
]

\saythanks
\section*{Introduction}
\label{sec:intro}
The XENON collaboration has recently reported an excess in the low-energy electron recoil data gathered during the Science Run 1~\cite{Aprile:2020tmw}. The signal concerns energies ranging from $1$ keV to $7$ keV, with a peak centered at about $2$ keV and a local statistical significance that exceeds the $3\sigma$ level. If interpreted in terms of new physics, the measurement seems to support the existence of a structured \textit{dark sector}~\cite{Takahashi:2020bpq, Kannike:2020agf, Lee:2020wmh, Choi:2020udy, Buch:2020mrg, AristizabalSierra:2020edu, Chen:2020gcl, Bell:2020bes, Du:2020ybt, Su:2020zny, Harigaya:2020ckz, Boehm:2020ltd, Fornal:2020npv,Alonso-Alvarez:2020cdv,Jho:2020sku,Baryakhtar:2020rwy,Bloch:2020uzh} as alternative (and simpler) solutions that propose solar axions, dark photons or large neutrino magnetic moments are generally disfavoured by stellar physics constraints~\cite{ Aprile:2020tmw,DiLuzio:2020jjp,An:2020bxd}.  Other explanations of the excess are considered in Refs.~\cite{Nakayama:2020ikz,Dey:2020sai,Paz:2020pbc, Primulando:2020rdk, Bally:2020yid} and the hypothesis that the signal was caused by a tritium background in the detector is not excluded~\cite{Aprile:2020tmw, Robinson:2020gfu}.

Insisting on explaining the {\tt XENON1T} excess with dark matter (DM), it is possible to obtain a good fit of the signal by considering a flux of fast moving particles that scatter on electrons~\cite{Kannike:2020agf} and, thereby, generate the observed spectrum. The span of required velocities, which in natural units is $v\sim0.03-0.25$ for DM masses $m\sim\mathcal{O}(10^{-4}-10)$ GeV, exceeds the virial velocity of cold dark matter (CDM), $v_{CDM}\sim\mathcal{O}(10^{-3})$, as well as the escape velocity of Milky Way ~\cite{Smith:2006ym}. This fast DM component is then likely generated locally, in our galaxy, and calls for a dark sector composed by at least two different particle species on top of new mediators that connect to the Standard Model (SM).

Interestingly, the scenario is aligned with the framework of boosted dark matter~\cite{Agashe:2014yua}, where a particle species $\psi$ plays the role of CDM and interacts with the SM only via gravity. The thermal relic abundance of this component is regulated by its interaction with a sub-dominant population of different particles, $\chi$, which instead couple to ordinary matter through new interactions. If the dark sector masses are such that $m_\psi\gtrsim m_\chi$, the $\chi$ particles produced in $\psi$ annihilation~\cite{Agashe:2014yua} or semi-annihilation~\cite{DEramo:2010keq} can have the velocities required to explain the {\tt XENON1T} signal~\cite{Fornal:2020npv,Kannike:2020agf}.

In the present Letter we consider the boosted DM solution of the {\tt XENON1T} measurement to analyze, in depth, the impact of radiative corrections. In fact, even if the CDM has no direct coupling to the SM, as the boosted DM component necessarily interacts with both $\psi$ particles and (at least) the electrons, it inevitably generates a coupling between these two species at the loop level. The whole CDM background will then contribute to the electron scattering signal, becoming observable~\cite{Aprile:2019xxb} if the relevant loop-suppressed cross section is compensated by the larger density of $\psi$: $\rho_\psi=\rho_{CDM}\simeq0.4$ GeV/cm$^3$\cite{Buch:2018qdr}. Similarly, the SM sources loop diagrams that connect the CDM component to the quarks via virtual electrons and photons, thereby yielding a potential signal in nuclear recoil searches~\cite{Aprile:2019xxb, Agnes:2018ves}. 

\section*{Simple models for two-component dark matter}

To investigate the mentioned effects in a manner as general as possible, we write down effective interactions that capture the gist of boosted DM models for fermion and scalar DM species. In particular, for the scalar scenario, we differentiate between semi-annihilation and annihilation as production mechanism for the boosted DM component. Henceforth, we denote the dominant and subdominant species with $\psi$ and $\chi$, respectively, regardless of the spin assignations.  

The effective Lagrangians we consider are:
\begin{itemize}
\item Scalar DM. 
\begin{align}
\label{eq:L0}
    \LG_{0}& \supset 
    (\pd_\mu \psi) (\pd^\mu \psi)^\dg - m_\psi^2 \psi^\dg\psi  
    \\ \nn &+ \frac12 (\pd_\mu \chi) (\pd^\mu \chi) 
   - \frac{m_\chi^2}{2} \chi^2 - \frac{\bar e e\chi^2}{\Lambda_e} 
   \\ \nn &
   - \lambda_s (\psi \psi \psi + \psi^\dg \psi^\dg \psi^\dg)  \chi 
   - \lambda_a \psi^\dg\psi \chi^2.
\end{align}
In the above equation, the coefficients $\lambda_s$ and $\lambda_a$ weight the production mechanisms of the subdominant DM component. The semi-annihilation can be singled out by imposing that the field $\psi$ transforms non-trivially under a $\mathbb{Z}_3$ symmetry\footnote{This motivates the use of a complex field $\psi$. }~\cite{DEramo:2010keq}. The interaction term between the electron, $e$, and DM involves the effective scale $\Lambda_e$ and is chosen so that the former interacts, at the tree level, only with $\chi$.  

\item Fermion DM.
\begin{align}
\label{eq:L12}
    \LG_{ 1/2} & \supset 
    \bar\psi (i\slashed{\pd}- m_\psi)\psi + \bar\chi (i\slashed{\pd}- m_\chi)\chi  
\\ \nn &
   - \frac{(\bar \psi \chi) (\bar \chi \psi)}{\Lambda_{a}^2}
   - \frac{(\bar e\gamma_\mu e)(\bar \chi\gamma^\mu \chi)}{\Lambda_{e}^2} .
\end{align}
\end{itemize}

\section*{Dark Matter abundance}

In order to simplify our analysis, we review the impact of CDM annihilation on the dark sector separately from the case of semi-annihilation, expecting that our results hold qualitatively also in scenarios where both the production mechanisms are allowed.  

\subsection*{Freeze-out via annihilation}

Considering pair annihilation processes first, the evolution of the CDM number density is regulated by the following Boltzmann equation~\cite{kolb1994early} 
\begin{equation}
    \frac{\td n_\psi}{\td t} + 3H n_\psi
    =
    -\sa (n_\psi^2 -{n^{eq}_\psi}^2), 
\end{equation}
where $H$ is the Hubble parameter and $n_\psi^{eq}$ the equilibrium number density of $\psi$ particles. The formula holds regardless of the chosen spin representation, which however affects the explicit expression for the annihilation cross section. The thermal average of the latter (times the relative velocity), $\sa$,  regulates the observed relic abundance of CDM according to
\begin{equation}
    \Omega_\psi \simeq 0.2 \left( \frac{5 \times 10^{-26} \, \text{cm}^3/ \text{s}}{\sa} \right),
\end{equation}
where we have used the $s$-channel approximation for $\sa $. Explicitly:
\begin{itemize}
    \item Scalar DM.
    \begin{equation}
        \sa \simeq \frac{\lambda_{a}^2}{8 \pi m_{\psi}^2}\sqrt{1 - \frac{m_{\chi}^2}{m_{\psi}^2}} . 
    \end{equation}
    \item Fermion DM.
    \begin{equation}
        \sa \simeq \frac{(m_{\chi}+ m_{\psi})^2 }{8 \pi \Lambda_a^4}\sqrt{1 - \frac{m_{\chi}^2}{m_{\psi}^2}} . 
    \end{equation}
\end{itemize}

\subsection*{Freeze-out via semi-annihilation}
For the semi-annihilation, the Boltzmann equation that tracks the $\psi$ number density is~\cite{DEramo:2010keq}
\begin{equation}
    \label{eq:Bpa}
    \frac{\td n_\psi}{\td t} + 3H n_\psi
    =
    -\ssa (n_\psi^2 -n_\psi n^{eq}_\psi) 
\end{equation}
and the relic abundance is determined by 
\begin{equation}
    \Omega_\psi \simeq 0.2 \left( \frac{15 \times 10^{-26} \, \text{cm}^3/ \text{s}}{\ssa} \right).
\end{equation}
Then, in the $s$-channel approximation, we obtain for the semi-annihilation cross section:
    \begin{equation}
        \ssa \simeq \frac{9 \lambda_s^2}{32 \pi m_{\psi}^2}\sqrt{9 - 10 \frac{m_{\chi}^2}{m_{\psi}^2} + \frac{m_{\chi}^4}{m_{\psi}^4}}.
    \end{equation}

\section*{Impact of loop-induced process}
In order to assess the impact of the mentioned loop-induced process, we start by fitting the {\tt XENON1T} result under the assumption that a flux of boosted $\chi$ particles produces the observed electron recoil. 

After accounting for the experimental resolution of the detector~\cite{Aprile:2020yad}, its efficiency~\cite{Aprile:2020tmw}, and considering the contribution of both the $4s$ and $3s$ orbitals to the transition (the latter dominates for energy deposits larger than the corresponding ionization energy of 1.17 KeV)~\cite{Roberts:2016xfw,Roberts:2019chv}, a profiled $\chi$-squared analysis reveals that the signal is well fit by $v_\chi/c \in [0.03, 0.25]$ and $m_\chi \in [ 10^{-4}, 10]$ GeV. The results we obtain agree with Ref.~\cite{Kannike:2020agf} and point to a value of the $\chi$ number density times the relevant cross section of $n_\chi \sigma_{\chi e} \simeq 10^{-43}$ cm$^{-1}$. 

Within the boosted DM framework, however, the number flux of the subdominant DM component is regulated by the same cross section that determines the CDM relic abundance. Assuming the NFW halo profile~\cite{Navarro:1995iw}, we then compute the full sky flux 
\begin{equation}
    \Phi_\chi = 1.6 \times 10^{-4} \, \text{cm}^{-2} / \text{s} \left( \frac{\braket{\sigma v}}{5 \times 10^{-26} \, \text{cm}^3/ \text{s} }   \right) \left(\frac{\text{GeV}}{m_\psi} \right)^2,
\end{equation}
and estimate the $\chi e$ cross section as
\begin{equation}
    \sigma_{\chi e}  \simeq  10^{-30}  \text{cm}^2 \left( \frac{5 \times 10^{-26} \, \text{cm}^3/ \textrm{s}}{\braket{\sigma v}} \right)   \left(\frac{m_\chi}{\text{GeV}} \right)^2 \left(  \frac{v_\chi/c}{0.1} \right),   
\end{equation}
where in both the equations $\braket{\sigma v} = \sa$ $(\ssa)$ for the case of annihilation (semi-annihilation).

We remark that the relatively large value obtained for the cross section justifies the assumption of a galactic origin of the $\chi$ background~\cite{Fornal:2020npv} and that considering different halo profiles can result in $\mathcal O(1)$ corrections to the above estimate which do not impact on our conclusions. Because the boost factor $\gamma_\chi = E_\chi / m_\chi$ resulting from annihilation and semi-annihilation are~\cite{Fornal:2020npv}
\begin{equation}
\gamma_\chi^{ann} = \frac{m_\psi}{m_\chi}, \qquad  \gamma_\chi^{sem} = \frac{5 m_\psi^2 - m_\chi^2}{4 m_\psi^2}   
\end{equation}
we also observe that the two DM components have comparable masses. 

\begin{figure}[ht]
\centering
    \includegraphics[height=3.5 cm]{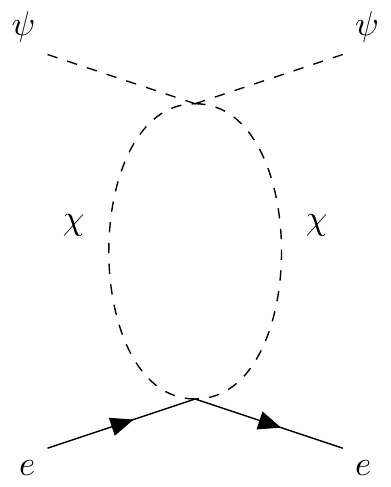}
    \hspace{.2cm}
    \includegraphics[height=3.5 cm]{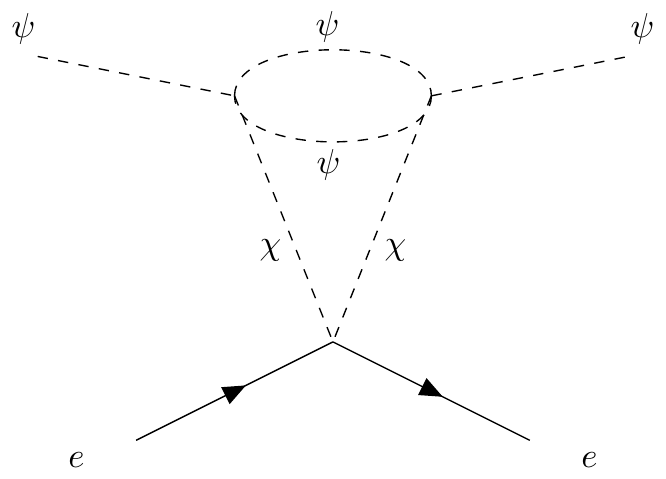}
    \caption{Loop-induced diagrams that connect the CDM component to the electrons for, respectively, the annihilation and semi-annihilation scenarios.}
    \label{fig:2}
\end{figure}

By using the effective Lagrangians in Eqs.~\eqref{eq:L0} and~\eqref{eq:L12}, we then compute the scattering cross section of CDM on electrons, sourced by the loop diagrams shown in Fig.~\ref{fig:2}. In terms of the $\chi e$ cross section we obtain in the leading-log approximation:
\begin{itemize}
    \item Annihilation, scalar DM.\\
    \begin{equation}
       \sigma_{\psi e} \simeq   \left[  \frac{\lambda_a^2}{256 \pi^4}   \log^2 \frac{m_\psi^2}{\mu^2}  \right]  \sigma_{\chi e},
    \end{equation}
    with $\sigma_{\chi e} = m_e^2/(4 \pi m_\chi^2 \, \Lambda_{e}^2)$.
    \item Annihilation, fermion DM.\\
    \begin{equation}
       \sigma_{\psi e} \simeq   \left[  \frac{9}{256 \pi^4}    \frac{m_\psi^4}{\Lambda_a^4} \log^2 \frac{m_\psi^2}{\mu^2}  \right]  \sigma_{\chi e},
    \end{equation}
    with $\sigma_{\chi e} = m_e^2/(\pi \Lambda_{e}^4)$.
    \item Semi-annihilation, scalar DM.\\
    \begin{equation}
        \sigma_{\psi e} \simeq 
        \lambda_s^4 \left[ \frac{9 \log^2(m_{\psi}^2/\mu^2)}{128 \pi^4} \right]^2  \sigma_{\chi e},
    \end{equation}
    with $\sigma_{\chi e} = m_e^2/(4 \pi m_\chi^2 \, \Lambda_{e}^2)$.
\end{itemize}
For the purpose of the numerical analysis, we set $\mu = \Lambda_e$ assuming that it is the lightest scale of new physics involved in the process.

\begin{figure}[ht]
    \centering
    \includegraphics[width=0.25 \textwidth]{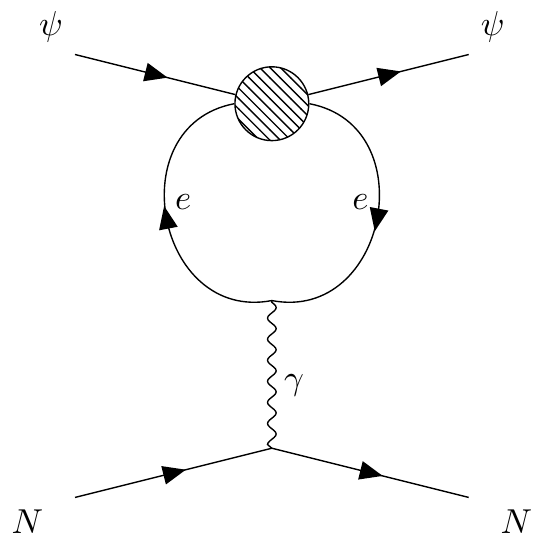}
    \caption{Radiative diagram that couples the CDM component to quarks, potentially leading to an observable nuclear recoil~\cite{PhysRevD.80.083502}.}
    \label{fig:3}
\end{figure}

The interaction with the electrons also induce a radiative coupling of $\psi$ to the quarks, which lead to potentially visible scattering of DM on the Xenon nucleons. In this case, the effective spin-independent nucleon cross section is sourced by the diagrams in Fig.~\ref{fig:3} and amounts to~\cite{PhysRevD.80.083502}:
\begin{equation}
    \sigma_{\psi N} \simeq \frac{\mu_N^2}{A^2} \left( \frac{\alpha Z}{3 \pi m_e} \right)^2 \log^2 \frac{m_e^2}{\Lambda_{e}^2}  \sigma_{\psi e}.
\end{equation}

\begin{figure}[ht]
    \includegraphics[width=0.45 \textwidth]{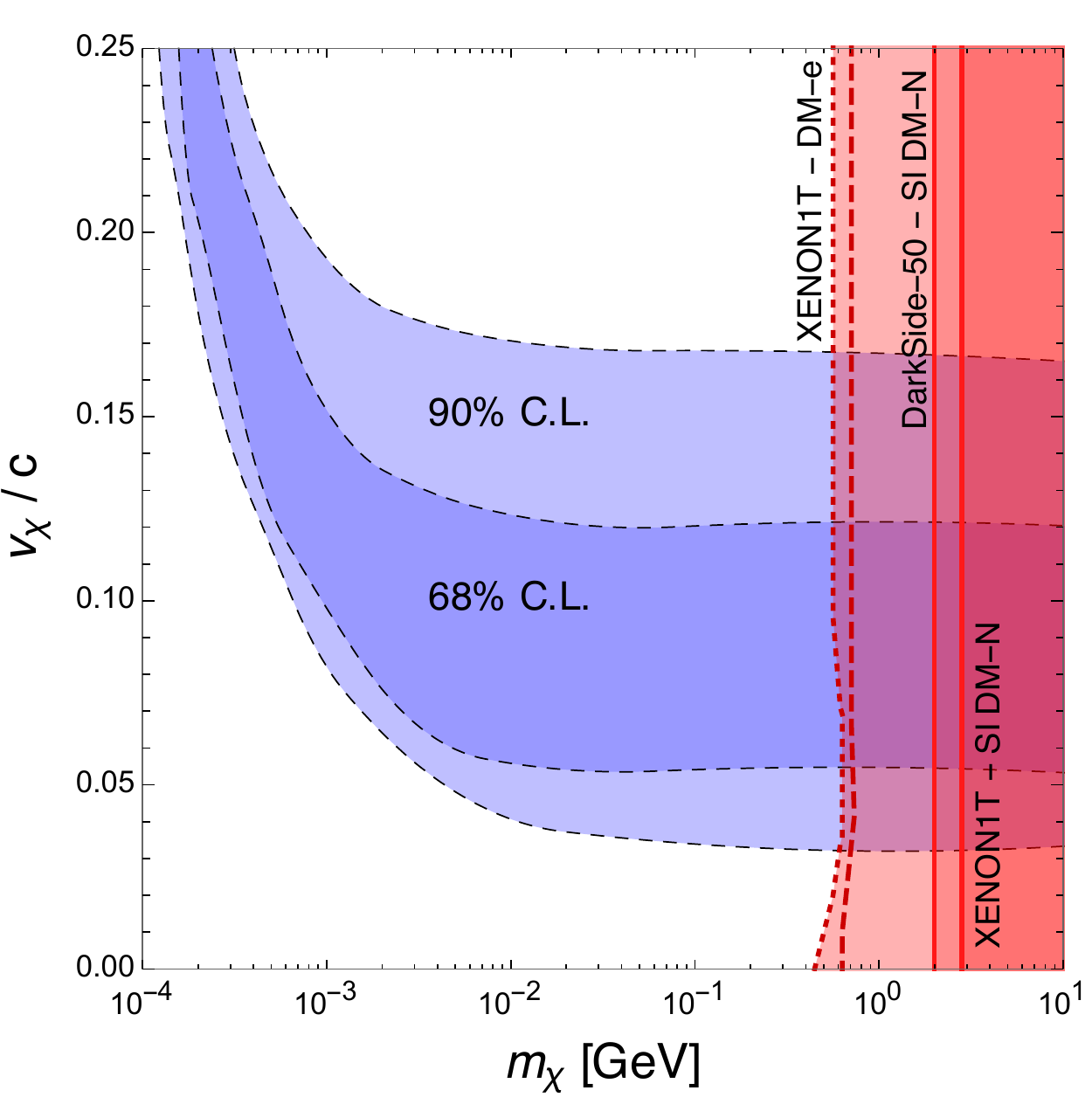}
    \caption{Impact of radiative corrections within the analyzed framework. The dark and light blue areas indicate the parameter space of boosted DM model selected by the {\tt XENON1T} excess at a 68\% and 90\% confidence level, respectively. The red bands show instead the exclusion contours due to the indicated searches, which probe the DM annihilation scenario through the analyzed loop-induced processes. The dotted limit bounds the scalar scenarios, whereas the dashed line is for the fermion one. The bounds due to spin-independent (SI) scattering on nucleons affect both the models. }
    \label{fig:1}
\end{figure}

The results obtained with the present analysis are summarized in Fig.~\ref{fig:1}. The dark (light) blue shade indicated the parameter space selected by fitting the {\tt XENON1T} signal within a confidence level of 68\% (95\%). As remarked before, the region agrees with the findings of Ref.~\cite{Kannike:2020agf}. The novelty of the present analysis is in the set of exclusion bounds that affect the upper range of DM masses. In correspondence of these regions, the radiative couplings considered in this work induce scattering cross sections that exceed the reported experimental bounds at a 90\% confidence level. The constraints shown in the plot apply only to the case of $\chi$ production via DM annihilation, with the dotted line denoting the electron recoil bound~\cite{Aprile:2019xxb} for the scalar scenario and the dashed one for the fermion case. The constraints due to spin-independent (SI) DM-nucleon scattering~\cite{Aprile:2019xxb, Agnes:2018ves}, indicated by the solid lines, apply to both the models.  

As for the case of semi-annihilation, not shown in the figure, we find that the mentioned constraints exclude only a small portion of the parameter space centered on $m_\chi=10$ GeV, regardless of the DM speed. The difference with respect to the annihilation scenario can be ascribed to the perturbative order at which the scattering processes are generated. Indeed, for semi-annihilations the interaction of $\psi$ with the electrons arises only at two-loop level.
The exclusion extends to lower masses if we allow for the CDM component to constitute only a fraction of the observed cosmological abundance, reaching $m_\chi \simeq 5$ GeV for $\Omega_{\psi}=0.02$.

\section*{Summary}
\label{sec:summary}

Motivated by the recent excess in electron nuclear recoils unveiled by the XENON collaboration, we have assessed the impact of loop-induced processes within the framework of boosted dark matter. 

The interactions of the sub-dominant dark matter component that, supposedly, induces the signal, inevitably couple the species acting as cold dark matter to the electrons at higher orders in perturbation theory. Similarly, quantum electrodynamics provides loop processes that make the cold dark matter component interact also with quarks. As a result, the cold dark matter background can leave a potentially observable imprints in the electron and nuclear recoils. 

The bounds we have obtained by computing the mentioned radiative processes are summarized in Fig.~\ref{fig:1}, and exclude the upper range of masses indicated by the signal fit regardless of the dark matter velocity. Interestingly, these constraints apply exclusively to the case where the boosted dark matter component is produced via the annihilation of the dominant species. In the case where semi-annihilation sources the flux responsible for the signal, the same processes yield only negligible bounds since arise at higher-order in the perturbative expansion. 

Regardless of the actual cause of the {\tt XENON1T} excess, our result is of value for the phenomenology of all boosted and multi-component DM scenarios that reduce to the analyzed effective models. 

\section*{Acknowledgements}
This work was supported by the Estonian Research Council grants PRG356, PRG803, MOBTT86, MOBTT5 and by the EU through the European Regional Development Fund
CoE program TK133 ``The Dark Side of the Universe". The authors thank Kristjan Kannike, Daniele Teresi and Hardi Veerm\"ae for useful discussion.  

\bibliographystyle{JHEP}
\bibliography{bib}

\end{document}